\newif\ifsingle
\singletrue 

\newif\ifproofs

\ifsingle
\documentclass[11pt,draftclsnofoot, onecolumn]{IEEEtran}		
\else		
\documentclass[10pt,final, twocolumn]{IEEEtran}
\fi

\usepackage{multirow} 
\usepackage{soul}
\usepackage{times}
\usepackage{amsmath,dsfont}
\usepackage{amssymb,amsthm}
\usepackage{epsfig,verbatim}
\usepackage{subfigure}
\usepackage{setspace}
\usepackage{color}
\usepackage{cite}
\usepackage{epstopdf}
\usepackage{graphics}
\usepackage{accents}
\usepackage{acronym}
\usepackage[bookmarks,colorlinks]{hyperref}
\usepackage{booktabs}
\usepackage{mathtools}
\usepackage{algorithm}
\usepackage{algorithmic}
\usepackage{enumitem}
\usepackage{tcolorbox}

\usepackage{multirow} 

\newcommand{\myVec}[1]{{\boldsymbol{#1}}}
\newcommand{\myMat}[1]{{\boldsymbol{#1}}}
\newcommand{\mySet}[1]{\mathcal{#1}}

\newcommand{\Revise}[1]{#1}

\definecolor{NewColor}{rgb}{0,0,0} 

\ifsingle

\else

\fi 

\definecolor{mypurple}{rgb}{0.910, 0.910, 0.969}
\definecolor{myblue}{rgb}{0.122, 0.435, 0.698}

\acrodef{adc}[ADC]{analog-to-digital convertor}
\acrodef{dac}[DAC]{digital-to-analog convertor}
\acrodef{cs}[CS]{compressed sensing}
\acrodef{dtft}[DTFT]{discrete-time Fourier transform}
\acrodef{bpsk}[BPSK]{binary phase shift keying}
\acrodef{ber}[BER]{bit error rate}
\acrodef{ofdm}[OFDM]{orthogonal frequency division multiplexing}
\acrodef{csi}[CSI]{channel state information}
\acrodef{map}[MAP]{maximum a-posteriori probability}
\acrodef{snr}[SNR]{signal-to-noise ratio}
\acrodef{sinr}[SINR]{signal-to-interference-plus-noise ratio}
\acrodef{bs}[BS]{base station} 
\acrodef{mimo}[MIMO]{multiple-input multiple-output}
\acrodef{mse}[MSE]{mean-squared error}
\acrodef{nmse}[NMSE]{normalized mean-squared error}
\acrodef{pdf}[PDF]{probability density function}
\acrodef{rv}[RV]{random variable}
\acrodef{lti}[LTI]{linear time-invariant}
\acrodef{wss}[WSS]{wide-sense stationary}
\acrodef{psd}[PSD]{power spectral density}
\acrodef{ser}[SER]{symbol error rate} 
\acrodef{isi}[ISI]{intersymbol interference} 
\acrodef{em}[EM]{expectation minimization} 
\acrodef{tdd}[TDD]{time division duplexing} 
\acrodef{ut}[UT]{user terminal} 
\acrodef{awgn}[AWGN]{additive white Gaussian noise}
\acrodef{cgac}[CGAC]{Complex-gain analog combiner}
\acrodef{psoac}[PSOAC]{Phase-shifter-only analog combiner}
\acrodef{fpga}[FPGA]{field-programmable gate array}
\acrodef{gui}[GUI]{grapichal user interface}
\acrodef{dfrc}[DFRC]{dual function radar-communications}
\acrodef{jrc}[JRC]{joint radar and communications}
\acrodef{pri}[PRI]{pulse repetition interval}
\acrodef{gsm}[GSM]{generalized spatial modulation}
\acrodef{smx}[SMX]{spatial multiplexing MIMO}
\acrodef{lfm}[LFM]{linear frequency modulation}
\acrodef{tws}[TWS]{track while scan}
\acrodef{mi}[MI]{mutual information}
\acrodef{fft}[FFT]{fast Fourier transform}
\acrodef{ifft}[IFFT]{inverse fast Fourier transform}
\acrodef{isi}[ISI]{inter symbol interference}
\acrodef{tdm}[TDM]{time division multiplexing}
\acrodef{fdm}[FDM]{frequency division multiplexing}
\acrodef{sm}[SM]{spatial modulation}
\acrodef{gsm}[GSM]{generalised spatial modulation}
\acrodef{sim}[SIM]{sub-carrier index modulation}
\acrodef{im}[IM]{index modulation}
\acrodef{rsaa}[RSAA]{randomized switched antenna array}
\acrodef{dsrc}[DSRC]{dedicated short range communications}
\acrodef{wlan}[WLAN]{wireless local area network}
\acrodef{mmwave}[mmWave]{millimeter-wave}
\acrodef{cdma}[CDMA]{code division multiple access}
\acrodef{emc}[EMC]{electromagnetic compatibility} 
\acrodef{rf}[RF]{radio frequency}
\acrodef{fmcw}[FMCW]{frequency-modulated continuous-wave}
\acrodef{far}[FAR]{frequency agile radar}
\acrodef{adas}[ADAS]{autonomous driving assistance system}
\acrodef{hrrp}[HRRP]{high range resolution profile}
\acrodef{cpi}[CPI]{coherent processing interval}
\acrodef{ula}[ULA]{uniform linear array}
\acrodef{simo}[SIMO]{single input multiple output}
\acrodef{summer}[SUMMeR]{sub-Nyquist MIMO radar}
\acrodef{fdma}[FDMA]{frequency-division multiple access}
\acrodef{lidar}[LIDAR]{light detection and ranging}
\acrodef{v2v}[V2V]{vehicle-to-vehicle} 
\acrodef{v2i}[V2I]{vehicle-to-infrastructure}
\acrodef{v2n}[V2N]{vehicle-to-network}
\acrodef{v2c}[V2C]{vehicle-to-cloud}
\acrodef{v2p}[V2P]{vehicle-to-pedestrian}
\acrodef{v2x}[V2X]{vehicle-to-everything}
\acrodef{cpm}[CPM]{continuous phase modulation}
\acrodef{stf}[STF]{short training field}
\acrodef{cef}[CEF]{channel estimation field}
\acrodef{ism}[ISM]{industrial scientific medical}
\acrodef{omp}[OMP]{orthogonal matching pursuit}
\acrodef{dft}[DFT]{discrete Fourier transform}
\acrodef{qpsk}[QPSK]{quadrature phase shift keying}

\IEEEoverridecommandlockouts

\title{Joint Radar-Communications Strategies for Autonomous Vehicles}

\author{
	\IEEEauthorblockN{ Dingyou Ma, Nir Shlezinger, Tianyao Huang,  Yimin Liu, and Yonina C. Eldar
	} 
	\thanks{D. Ma, T. Huang and Y. Liu are with the Department of Electronic Engineering, Tsinghua University, Beijing, China (e-mail: mdy16@mails.tsinghua.edu.cn; \{huangtianyao, yiminliu\}@tsinghua.edu.cn).
		N. Shlezinger  and Y. C. Eldar are with the Faculty of Mathematics and Computer Science, Weizmann Institute of Science, Rehovot, Israel (e-mail: nirshlezinger1@gmail.com; yonina.eldar@weizmann.ac.il). 
		This work received funding from the National Natural Science Foundation of China under grants 61801258 and 61571260, from the European Union’s Horizon 2020 research and innovation program under grant No. 646804-ERC-COG-BNYQ, and from the Air Force Office of Scientific Research under grant No. FA9550-18-1-0208.
	}
	\vspace{-1.2cm}
}
\vspace{-0.75cm}

\begin{document}
	
	\maketitle
	\pagestyle{plain}
	\thispagestyle{plain}
	\begin{abstract} 
		Self-driving cars constantly asses their environment in order to choose routes, comply with traffic regulations, and avoid hazards. To that aim, such vehicles are equipped with wireless communications transceivers as well as multiple sensors, including automotive radars. The fact that autonomous vehicles implement both radar and communications motivates designing these functionalities in a joint manner. Such dual function radar-communications (DFRC) designs are the focus of a large body of recent works. These approaches can lead to substantial gains in size, cost, power consumption, robustness, and performance, especially when both radar and communications operate in the same range, which is the case in vehicular applications. This article surveys the broad range of DFRC strategies and their relevance to autonomous vehicles. We identify the unique characteristics of automotive radar technologies and their combination with wireless communications requirements of self-driving cars. 
		Then, we map the existing DFRC methods along with their pros and cons in the context of autonomous vehicles, and discuss the main challenges and possible research directions for realizing their full potential. 
	\end{abstract}

	\vspace{-0.4cm}
	\section{Introduction} 
	\label{sec:Intro}
	\vspace{-0.1cm}
	Autonomous vehicles are required to navigate efficiently and safely in a wide variety of complex uncontrolled environments. To meet these requirements, such self-driving cars must be able to reliably sense and interact with their surroundings.  This acquired sensory information as well as data communicated from neighboring vehicles and road-side units are essential to avoid obstacles, select routes, detect hazards, and comply with traffic regulations, all in real-time.

	In order to reliably sense the environment, autonomous vehicles are equipped with multiple sensing technologies, including computer vision acquisition, i.e., cameras, \ac{lidar} laser-based sensors, global positioning systems, and radar transceivers. 
	Each of these technologies has its advantages and disadvantages. In order to allow accurate sensing in a broad range of complex environments, self-driving cars should simultaneously utilize all of these aforementioned sensors. Radar for instance, provides the ability to accurately detect distant objects and is typically more robust to weather conditions and poor visibility compared to its competing sensing technologies \cite{Patole2017}.
	
	Radar systems, which detect the presence of distant objects by measuring the reflections of electromagnetic probing waves, have been in use for over a century. Radar has been most commonly used in military applications, aircraft surveillance, and navigation systems. The application of radar for vehicles, referred to as automotive radar \cite{meinel2014evolving}, is substantially different from traditional radar systems: Most notably, automotive radar systems, which are used by mass-produced vehicles, are far more limited in size, power, and cost. Furthermore, while conventional radar aims to detect a relatively small number of distant targets, e.g., airplanes, automotive radar is required to sense in complex dense urban environments in which a multitude of scatterers at close ranges should be accurately detected. Despite these differences, automotive radar is an established and common technology nowadays, and the vast majority of newly manufactured vehicles are equipped with radar-based \acp{adas} \cite{Patole2017}.   
	
	In addition to their ability to sense their environment, autonomous vehicles are also required to carry out various forms of communications, as illustrated in Fig. \ref{fig:Commlinks}:  \ac{v2v} transmissions allow self-driving cars to share their attributes with neighbouring vehicles; \ac{v2i} messages facilitate intelligent road management by conveying information between cars and road-side units; 
	\ac{v2p} communications can be used to warn or alarm nearby pedestrians; Finally, service providers and cloud applications exchange possibly large amounts of data with self-driving cars via \ac{v2n} and \ac{v2c} links, respectively. The resulting broad range of different   tasks, which substantially vary in their latency, throughput, and reliability requirements, can be implemented by using individual communications technologies for each application, or by using a unified \ac{v2x} strategy \cite{wang2017overview}, possibly building upon the cellular infrastructure.
	
	\ifsingle	
	\begin{figure}	
		\centerline{\includegraphics[width=0.6\columnwidth] {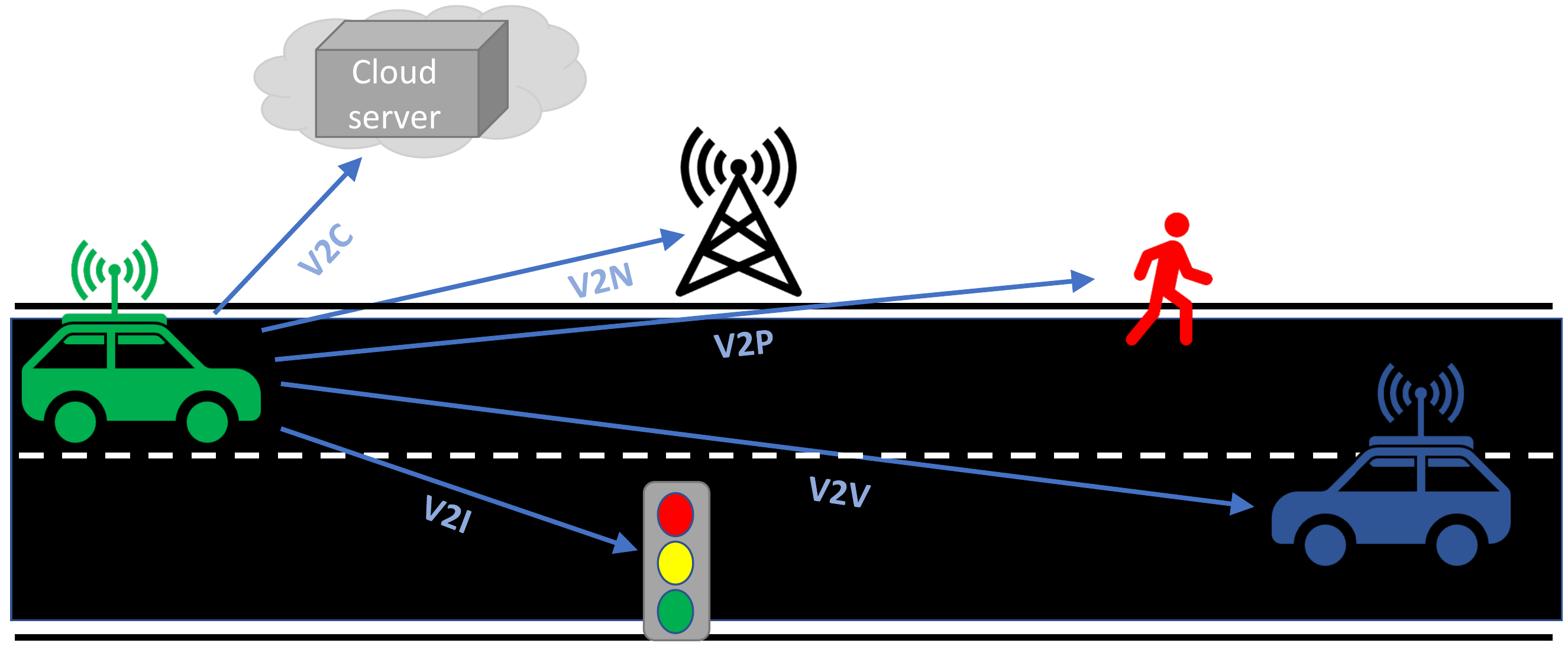}}
		\vspace{-0.2cm}
		\caption{Autonomous vehicles communications links.}
		\label{fig:Commlinks}
	\end{figure} 
	\else
	\begin{figure}	
		\centerline{\includegraphics[width=\columnwidth] {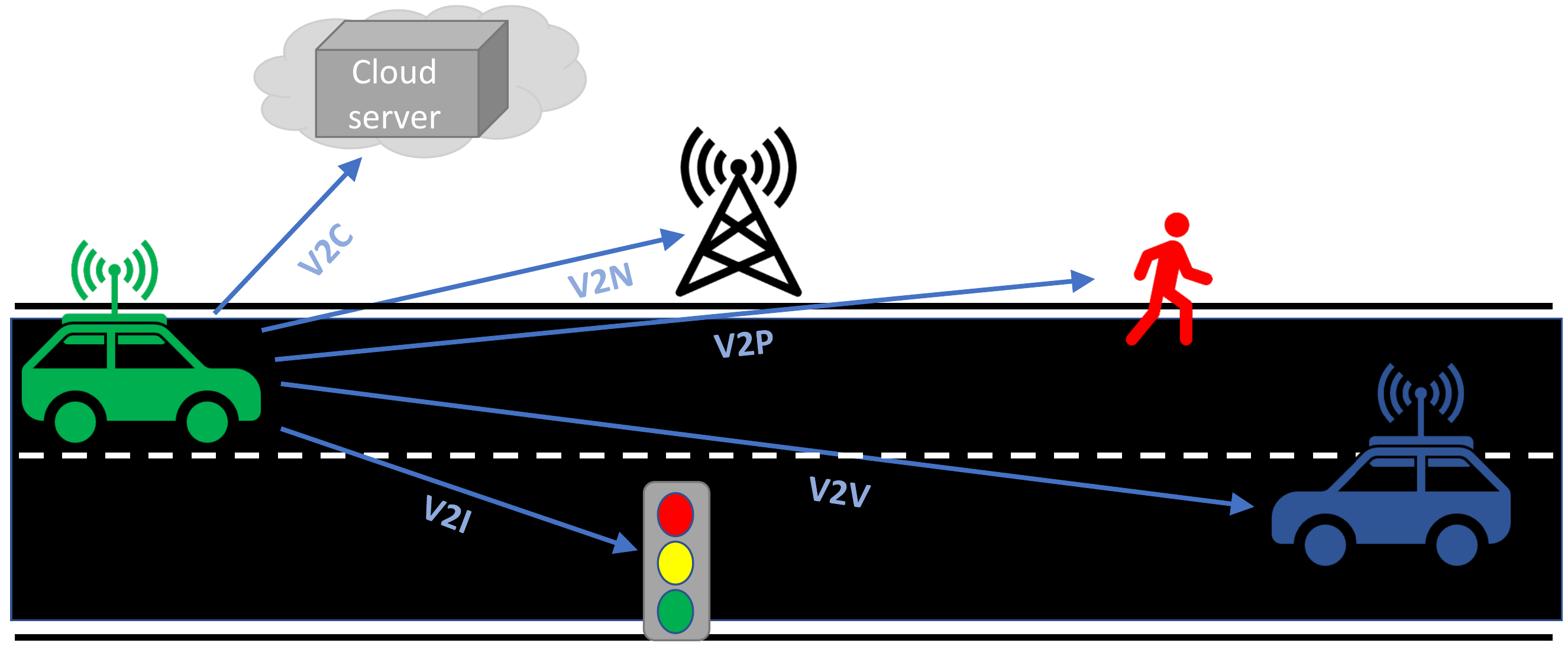}}
		\vspace{-0.2cm}
		\caption{Autonomous vehicles communications links.}
		\label{fig:Commlinks}
	\end{figure} 
	\fi
	
	Automated cars thus implement two technologies which rely on the transmission and processing of electromagnetic signals: radar and wireless communications.
	A possible approach in designing self-driving cars is to use individual systems for radar and communications, each operating separately. An alternative strategy is to {\em jointly design} these functionalities as a \ac{dfrc} system. Such  schemes are the focus of extensive recent research attention \cite{Paul2017, mishra2019towards,gameiro2018research, han2013joint,Hassanien2016Signal, Liu2017Adaptive,Hassanien2016Dual,Hassanien2016Non,Sturm2011Waveform, McCormick2017Simultaneous,Liu2018MUMIMO, Liu2018Toward, Zheng2019Radar, Bica2019Multicarrier,  Ma2019A, Huang2019A,Huang2019multi,Reichardt2012Demonstating,Kumari2015Investigating,Kumari2018IEEE80211ad, Chiriyath2017,aydogdu2019radchat}. In particular, it was shown that jointly implementing radar and communications contributes to reducing the number of antennas \cite{Tavik2005Advanced},  system size, weight, and power consumption \cite{Liu2017Adaptive}, as well as alleviating concerns for \acl{emc} and spectrum congestion \cite{Hassanien2016Signal}. \Revise{Utilizing such joint designs in vehicular systems can mitigate the mutual interference among neighboring cars, facilitate coordination, and improve  pedestrian detection \cite{aydogdu2018improved}.\label{txt:Intro_Pedestrian}} These benefits make \ac{dfrc} systems an attractive technology for autonomous vehicles. 
	
	While the conceptual advantages of joint radar-communications designs for autonomous vehicles are clear, the proliferation of different \ac{dfrc} strategies makes it difficult to identify what scheme is most suitable for which scenario. For example, some \ac{dfrc} methods use existing \ac{v2x} communications waveforms as radar probing signals, thus allowing high communication throughput with relatively limited sensing capabilities \cite{Reichardt2012Demonstating,Kumari2015Investigating,Kumari2018IEEE80211ad}. Alternative schemes embed digital messages in the radar probing signals \cite{Huang2019multi,Huang2019A}, thus supporting low data rates which may be more suitable to serve as an additional channel to the standard communications functionalities of autonomous vehicles.  
	
	The goal of this article is to review \ac{dfrc} technologies in light of the unique requirements and constraints of self-driving cars, facilitating the identification of the proper technology for different scenarios. We begin by reviewing the basics of automotive radar, identifying its main challenges, recent advances, and fundamental differences from conventional radar systems. We then survey \ac{dfrc} methods, dividing previously proposed approaches into  \Revise{four} main categories: \Revise{coordinated} signals transmission methods utilizing individual signals for each functionality; communications waveform-based schemes, which use the communications signal as a radar probing waveform;  
	radar waveform-based techniques, which embed the digital message into the parameters of the radar signal; 
	\Revise{and the design of dedicated dual-function waveforms. We detail a representative set of \ac{dfrc} methods for each category}, and provide a map of the existing strategies in terms of their radar capabilities, information rates, and complexity.

	\vspace{-0.2cm}
	\section{Basics of Automotive Radar}
	\label{sec:AutomoRadar}
	\vspace{-0.1cm}
	Past decades have witnessed growing interest in automotive radar to improve the safety and comfort of drivers. A typical \ac{adas} implements various radar subsystems that enable functions including adaptive cruise control, blind spot detection,  and parking assistance\cite{Patole2017}. To understand the benefits of combining automotive radar with digital communications, we first review the basics of automotive radar.

	Automotive radars operate under different requirements and constraints compared to conventional radars, such as those utilized in military applications and air traffic control. 	First, conventional radar systems are required to detect a relatively small number of targets in ranges on the order of tens or hundreds of kilometers, while automotive radars must detect a multitude of objects in short ranges on the order of a few tens of meters. Furthermore, automotive radars are   incorporated into mass-produced vehicles, hence have more strict constraints on cost, size,   power consumption, and spectral efficiency compared to conventional radar. 
	Finally, automotive radars are   densely deployed urban environments, thus must be robust to interference while inducing minimal interference to neighboring radar systems. 
	
	Various techniques have been proposed to overcome the aforementioned challenges. In Table \ref{tbl:radar} we summarize the main challenges along with the leading methods to tackle them. It is noted that no single radar scheme is suitable to handle the complete set of requirements. For example, the  popular \ac{fmcw} waveform (see \emph{\ac{fmcw} Radar} on Page \pageref{Box:fmcw}), which can be operated using simplified hardware components, suffers from high sensitivity to interference; \ac{ofdm}  radar (described in \emph{\ac{ofdm} Waveform Radar} on Page \pageref{Box:ofdmradar}), which is suitable for multiuser scenarios, tends to require relatively costly hardware compared to alternative radars.
	\begin{table*}[t]
		\centering
		\caption{Automotive radar requirements.} 
		\vspace{-0.2cm}
		\label{tbl:radar}
		\ifsingle
		\small
		\fi 
		\begin{tabular}{|p{3.8cm}|p{11.7cm}|}
			\hline
			Requirements & Possible solutions    \\ \hline\hline 
			Operating in short ranges & Utilize separate transmit and receive antennas to process short range echoes.\\
			\hline
			\multirow{2}{*}{Limited antenna size}& Operate at \ac{mmwave} bands  using patch antennas.\\  
			&Increase virtual aperture (see \emph{\acs{mimo} Radar} on Page \pageref{Box:mimoradar}). \\  
			\hline 
			\multirow{2}{*}{Simplified hardware} &  Constant envelope signalling. \\ &Low-complexity de-chirp recovery, e.g., \ac{fmcw} (see \emph{\ac{fmcw} Radar} on Page \pageref{Box:fmcw}). \\
			\hline
			
			\multirow{1}{*}{Low power amplifiers}& Continuous or high duty cycle waveform, e.g., \ac{fmcw}.  \\  
			\hline 
			
			\multirow{2}{*}{Interference robustness}& Divide spectrum using \ac{ofdm}  (see \emph{\ac{ofdm} Waveform Radar} on Page \pageref{Box:ofdmradar}).\\ 
			& \Revise{Introduce}  agility (see \emph{Frequency Agile Radar} on Page \pageref{Box:far}) to increase survivability. \\  
			\hline		
		\end{tabular}  
	\end{table*}
	An additional aspect which should be considered in selecting an automotive radar scheme is its capability to be combined with wireless communications. 
	The fact that self-driving cars utilize both  radar and digital communications  motivates their joint design as a \ac{dfrc} system, as discussed in the following section. 
	
	\begin{tcolorbox}[float,
		toprule = 0mm,
		bottomrule = 0mm,
		leftrule = 0mm,
		rightrule = 0mm,
		arc = 0mm,
		colframe = myblue,
		colback = mypurple,
		fonttitle = \sffamily\bfseries\large,
		title =\ac{fmcw} Radar]	
		
		\ac{fmcw} is a continuous constant modulus radar waveform with a linearly modulated frequency, which can be generated and detected using simplified hardware.  To present \ac{fmcw}, we consider a radar system equipped with a single transmit antenna and a \ac{ula} with $L_R$ elements for receiving. In each radar \ac{cpi}, $M$ \ac{fmcw} pulses of duration $T_p$ are periodically transmitted with a \ac{pri} denoted by $T_{PRI}$, where $T_{PRI}$ is slightly larger than $T_p$. The $m$th pulse is given by $s_{m}(t) = e^{j2\pi f_c t + j \pi \gamma \left(t - mT_{PRI}\right)^2 }$, $t \in [m T_{PRI}, mT_{PRI} + T_p]$,
		where $f_c$ is the carrier frequency, and $\gamma$ is the frequency modulation rate. 
		
		\ \ To formulate the received signal, assume $P$ targets are located in the far field. The distance, velocity and angle of the $p$th target are denoted as $r_p$, $v_p$ and $\theta_p$, respectively. For the $p$th target, with the far field assumption, the round time delay between the transmit antenna and the $l$th receiver is $\tau_{l,p} = \frac{2\left(r_p +v_p t\right) - ld\sin\theta_p}{c}$,
		where $d$ is the distance between adjacent elements in the receiving array and $c$ is the speed of light.
		The radar echo received in the $l$th receiving antenna during the $m$th transmit pulse is represented as $	r_{m,l}\left(t\right) = \sum_{p=0}^{P-1}\alpha_{p}s_{m}\left(t- \tau_{l,p}\right) + w\left(t\right)$,
		where $\alpha_{p}$ is the complex reflective factor of the $p$th target and $w\left(t\right)$ is  additive white Gaussian noise. 
		
		\ \ To process the received signal,  $r_{m,l}\left(t\right)$ is mixed with the transmit signal. This procedure, referred to as de-chirp, yields a demodulated signal  given by $	y_{m,l}\left(t\right) = r_{m,l}\left(t\right)\cdot s_{m}^{\ast}\left(t\right)$, i.e.,
		\ifsingle
		\vspace{-0.2cm}
		\begin{align}
		y_{m,l}\left(t\right) 
		& = \sum_{p=0}^{P-1} \tilde{\alpha}_p e^{-j2\pi\left(\frac{2\gamma r_p}{c} + \frac{2v_pf_c}{c}\right)\left(t- mT_{PRI}\right)- j2\pi f_{c} \frac{2mv_{p}T_{PRI}}{c} + j2\pi\frac{ldf_c \sin\theta_p}{c}} + \tilde{w}\left(t\right),
		\vspace{-0.2cm}
		\label{eqn:FMCWrx}
		\end{align}
		\else
		\begin{align}
		&y_{m,l}\left(t\right)   = \sum_{p=0}^{P-1} \tilde{\alpha}_p e^{-j2\pi\left(\frac{2\gamma r_p}{c} + \frac{2v_pf_c}{c}\right)\left(t-mT_{PRI}\right)} \notag \\
		&\times e^{ - j2\pi f_{c} \frac{2mv_{p}T_{PRI}}{c} + j2\pi\frac{ldf_c \sin\theta_p}{c}}  + \tilde{w}\left(t\right),
		\vspace{-0.2cm}
		\label{eqn:FMCWrx}
		\end{align}
		\fi		
		where $(\cdot)^{\ast}$ is the complex conjugate, $\tilde{\alpha}_p := \alpha_{p}e^{-j2\pi f_c\frac{2r_p}{c}}$, and $\tilde{w}\left(t\right) := w\left(t\right)\cdot s_{m}^{\ast}\left(t\right)$.
		
		\ \ After de-chirp, the waveform frequency is typically much smaller compared to the bandwidth of the transmitted waveform, and can  be sampled with low speed \acp{adc}. It follows from \eqref{eqn:FMCWrx} that the targets range, velocity, and direction can then be recovered from the 3-dimensional \ac{dft} of the sampled $	y_{m,l}$, in the fast time domain (within a pulse), slow time domain (between  pulses) and spatial domain (over  antennas), respectively.
		
		\label{Box:fmcw}
	\end{tcolorbox}

	\vspace{-0.2cm}
	\section{Overview of Dual Function Systems}
	\label{sec:OverviewJRC}
\begin{figure}	
		\centerline{\includegraphics[width=0.9\columnwidth] {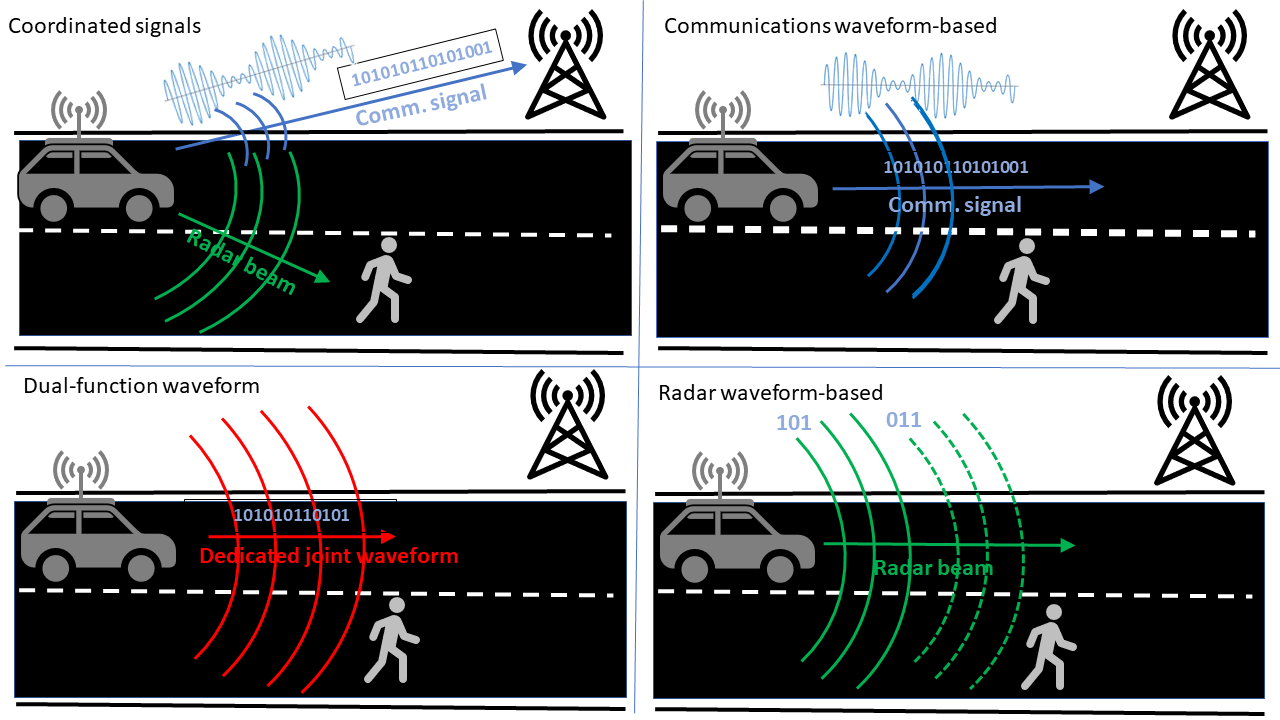}}
		\vspace{-0.2cm}
		\caption{An illustration of \ac{dfrc} strategies for autonomous vehicles. \Revise{Blue, green, and red waveforms represent communications signals, radar beams, and dedicated dual-function waveforms, respectively.}}
		\label{fig:DFRCCompare}
	\end{figure}

	\vspace{-0.1cm}
	Since \ac{dfrc} systems implement both radar and communications using a single  device, these functionalities inherently share some of the system resources, such as spectrum, antennas, and power. 
	Broadly speaking, existing \ac{dfrc} methods can be divided into \Revise{four} main categories as illustrated in Fig.~\ref{fig:DFRCCompare}: \Revise{coordinated separated signals transmission}, communications waveform-based approaches,  radar waveform-based schemes, \Revise{and joint dual-function waveform designs}. In the following, we review each of these categories, and discuss their pros and cons in the context of autonomous vehicles.  Throughout this section, we consider a \ac{dfrc} system jointly implementing a radar transceiver as well as the transmission of digital messages using $L_T$ transmit antennas (for both radar and communications) and $L_R$ receive antennas (for radar). For simplicity, we assume a single  communications receiver equipped with a single antenna.

	\vspace{-0.2cm}
	\subsection{Separate Coordinated Signals}
	\label{subsec:jrc_Coexistence}
	\vspace{-0.1cm}
	A common \ac{dfrc} approach is to utilize different signals for radar and communications, designing the functionalities to mitigate their cross interference, as illustrated in the upper-left subfigure in Fig. \ref{fig:DFRCCompare}. 
	Here, the $L_T \times 1$ transmitted signal can be written as 
	\vspace{-0.2cm}
	\begin{equation}
	\label{eqn:coexist}
	\myVec{s}(t) = \myVec{s}^{(r)}(t) + \myVec{s}^{(c)}(t),
	\vspace{-0.2cm}
	\end{equation}
	where $ \myVec{s}^{(r)}(t)$ is the radar probing waveform, and $\myVec{s}^{(c)}(t)$ is the continuous-time communications signal. 
	The ability to jointly transmit two dedicated signals with limited cross interference is typically achieved using either orthogonality boosting by division in time and/or frequency, or via spatial beamforming.
	
	\subsubsection{Time/Frequency Division}
	Arguably the most simple method to mitigate cross interference is to allocate a different frequency band to each waveform, commonly dictated by regulated spectrum allocation, or, alternatively,  a different time slot.  In such cases, the signals $\myVec{s}^{(r)}(t)$ and $\myVec{s}^{(c)}(t)$ in \eqref{eqn:coexist} either reside in different bands (for frequency division), or satisfy $\myVec{s}^{(r)}(t)\big(\myVec{s}^{(c)}(t)\big)^T = \myMat{0}$ at each time instance (for time division).
	Since system resources are allocated between both subsystems, these strategies inevitably result in a trade-off between radar and communications performance \cite{Chiriyath2017}.
	
	A straight-forward approach is to allocate the resources in a fixed or arbitrary manner. For instance, in \cite{Tavik2005Advanced}, a \ac{dfrc} system is achieved by using fixed non-overlapping bands and antennas. 
	A random antenna  allocation scheme is proposed in \cite{Ma2019A}, jointly enhancing the radar angular resolution and the communication rates. \Revise{The work \cite{aydogdu2019radchat} proposed a media access protocol for automotive \ac{dfrc} systems with time and frequency division to mitigate  interference with neighboring radars.\label{txt:Sepa_MtiInterfere}} These approaches assume that each functionality has its own frequency band. Using \ac{ofdm} signaling, i.e., letting the entries of $\myVec{s}^{(r)}(t)$ and $\myVec{s}^{(c)}(t)$  represent \ac{ofdm} radar and communications waveforms  (see \emph{\ac{ofdm} Waveform Radar} on Page \pageref{Box:ofdmradar}), respectively, allows to divide the spectrum in an optimized manner, as we detail next.
	
	Consider a frequency band divided into $N$ subbands. The discrete-time transmitted signal from the $l$th transmit antenna can be written as the $N \times 1$ vector $\myVec{s}_l$. Since the spectrum is divided into radar and communications, $\myVec{s}_l$ is given by
	\vspace{-0.2cm}
	\begin{equation}
	\label{equ:OFDMcoexist}	
	\myVec{s}_l = \myMat{F}^{H}\left[\myMat{U}_l\myVec{s}_l^{(r)} +\left(\myMat{I} - \myMat{U}_l\right)\myVec{s}_l^{(c)}\right],
	\vspace{-0.2cm}
	\end{equation}
	where $\myMat{F}^{H}$ is the inverse discrete Fourier transform matrix; the $N \times 1$ vectors $\myVec{s}_l^{(r)}$  and $\myVec{s}_l^{(c)}$ denote the \ac{ofdm} radar and communications symbols, respectively, in the frequency domain; and  $\myMat{U}_l$ is a diagonal matrix of size $N\times N$ with elements 0 or 1, representing the subcarrier selection at the $l$th element. 
	
	Setting the matrix $\myMat{U}_l$ in \eqref{equ:OFDMcoexist} determines how the bandwidth is divided. The work \cite{Sturm2013Spectrally} showed that when $\myMat{U}_l$ represents spectral interleaving, i.e., the support of its diagonal consists of multiple bulks of zeros and ones, radar resolution is comparable to that using the complete spectrum. 
	When the \ac{dfrc} system has a-priori knowledge of the statistical model of the radar target response and the communications channel, the subcarrier selection matrix $\myMat{U}_l$ can be set to optimize a linear combination of the radar target-echo mutual information and the communications input-output mutual information, as proposed in \cite{Bica2019Multicarrier}.

	\subsubsection{Spatial Beamforming}
	The utilization of multiple antennas enables to mitigate the mutual interference through spatial beamforming, for example, by projecting the radar waveform into the null space of its channel to the communications receiver \cite{Sodagari2012AProjection}, resulting in a zero forcing beamformer. 
	While such beamforming was originally proposed for separate systems, it can also be utilized for a \ac{dfrc} system. 
	

	\Revise{In this model, the communications and radar signals are beamformed using the matrices $\myMat{U}^{\left(c\right)}$ and $\myMat{U}^{\left(r\right)}$, respectively,  in order to mitigate the mutual interference while satisfying the performance constraints. The signals received at the communications receiver and  the radar target with direction $\theta$ are thus 
		\ifsingle		
		\vspace{-0.2cm}
		\begin{align}
		y^{\left(c\right)} = \myVec{h}^{T}\left(\myMat{U}^{\left(c\right)}\myVec{s}^{\left(c\right)} +  \myMat{U}^{\left(r\right)}\myVec{s}^{\left(r\right)}\right) + w^{\left(c\right)}, \ \mathrm{and } \ 
		y_{\theta}^{\left(r\right)} = \myVec{a}^{T}\left(\theta\right)\left(\myMat{U}^{\left(c\right)}\myVec{s}^{\left(c\right)} + \myMat{U}^{\left(r\right)}\myVec{s}^{\left(r\right)}\right),
		\label{equ:BeamForm1}
		\vspace{-0.2cm}
		\end{align}
		\else
		\begin{align}
		y^{\left(c\right)} &= \myVec{h}^{T}\left(\myMat{U}^{\left(c\right)}\myVec{s}^{\left(c\right)} +  \myMat{U}^{\left(r\right)}\myVec{s}^{\left(r\right)}\right) + w^{\left(c\right)}; \notag \\ 
		y_{\theta}^{\left(r\right)} &= \myVec{a}^{T}\left(\theta\right)\left(\myMat{U}^{\left(c\right)}\myVec{s}^{\left(c\right)} + \myMat{U}^{\left(r\right)}\myVec{s}^{\left(r\right)}\right),
		\label{equ:BeamForm1}
		\vspace{-0.2cm}
		\end{align}
		\fi
		where $\myVec{h}$ is the channel response from the \ac{dfrc} transmitter to the communications receiver, and $\myVec{a}\left(\theta\right)$ is the steering vector of the \ac{dfrc} transmitter to the radar target in direction $\theta$. Using the formulation \eqref{equ:BeamForm1}, the beamforming matrices $\myMat{U}^{\left(c\right)}$ and $\myMat{U}^{\left(r\right)}$ are jointly designed to mitigate cross interference while satisfying the performance requirements, e.g., maximize the \ac{sinr} at the communications receiver while meeting a given radar beampattern \cite{liu2019joint}.}
	
	
	\smallskip
	A clear advantage of the separated signals transmission strategy is that it can provide a wide variety of possible performance combinations. For time/frequency division schemes, the performance is determined by how the system resources, such as spectrum and time slots, are allocated to each functionality. 
	The performance trade-offs may be potentially improved using spatial beamforming,  allowing each functionality to utilize the full bandwidth and operate simultaneously at all time slots. 
	However,  the spatial beamformer is designed based on a-priori channel knowledge, which may be unavailable for fast moving vehicles. 
	According to the discussions above, time/frequency division-based schemes are likely to be more attractive in automotive applications. Since properly optimizing the resource allocation to achieve a desired performance trade-off requires considerable computation, fixed sub-optimal allocations, such as spectral interleaving, may be preferable.
	
	
	\begin{tcolorbox}[float,
		toprule = 0mm,
		bottomrule = 0mm,
		leftrule = 0mm,
		rightrule = 0mm,
		arc = 0mm,
		colframe = myblue,
		colback = mypurple,
		fonttitle = \sffamily\bfseries\large,
		title =\ac{ofdm} Waveform Radar]	
		For an \ac{ofdm} waveform radar of $M$ pulses with $N$ subcarriers, the transmit signal at the $m$th pulse is 
		\ifsingle		
		\vspace{-0.1cm} 
		\begin{equation}
		s_{m}\left(t\right) = \sum_{n=0}^{N-1}a_{m,n}{\rm rect}\left(\frac{t-mT_{\rm O}}{T_{\rm O}}\right)e^{j2\pi\left(f_c + f_n\right)t}.
		\label{eqn:OFDM1}
		\vspace{-0.1cm}
		\end{equation}
		\else
		\begin{equation}
		\!\!	s\!\left(t\right) \!=\! \sum_{m=1}^{M}\sum_{n=1}^{N}\!a_{m,n}{\rm rect}\left(\frac{t\!-\!mT_{\rm O}}{T_{\rm O}}\right)e^{j2\pi\left(f_c\! + \!f_n\right)t}.
		\label{eqn:OFDM1}
		\vspace{-0.1cm}
		\end{equation}
		\fi		
		In \eqref{eqn:OFDM1}, $\{a_{m,n}\} \in \mySet{A}$ are complex weights transmitted over the $m$th symbol on carrier $f_n$, which can be either fixed or randomized from some discrete set $\mySet{A}$;  ${\rm rect}\left(t\right)$ is a rectangular window of  unity support, $T_{\rm O} = T_{\rm S} + T_{\rm CP}$ is the \ac{ofdm} symbol duration; $T_{\rm S}$ is the elementary symbol duration; $T_{\rm CP}$ is the duration of the cyclic prefix; and $f_n = \frac{n}{T_{\rm S}}$. Using the notations in \eqref{eqn:FMCWrx}, the radar echo received in the $l$th receiving antenna is from $P$ targets, and can be approximated by
		\textcolor{black}{
			\ifsingle			
			\vspace{-0.1cm}
			\begin{align}
			y_{m,l}\left(t\right) \!\approx\! \sum_{p=0}^{P-1}\!\sum_{n=0}^{N-1}\tilde{\alpha}_{p} \!\cdot \!a_{m,n} {\rm rect}\!\left(\!\frac{t-mT_{\rm O} - \frac{2r_p}{c}}{T_{\rm O}}\!\right)\! e^{j2\pi f_n\left(t - \frac{2r_p}{c}\right) - j 2\pi f_c \frac{2v_p}{c}t + j2\pi \frac{ldf_{c}\sin\theta_p}{c}} \!+ \! w\left(t\right).
			\vspace{-0.1cm}
			\label{eqn:OFDM2}
			\end{align}
			\else
			\begin{align}
			&y_{m,l}\left(t\right) \approx \sum_{p=0}^{P-1}\sum_{n=0}^{N-1}\tilde{\alpha}_{p} \cdot a_{m,n} {\rm rect}\left(\frac{t-mT_{\rm O} - \frac{2r_p}{c}}{T_{\rm O}}\right)\notag \\
			&\times e^{j2\pi f_n\left(t - \frac{2r_p}{c}\right) - j 2\pi f_c \frac{2v_p}{c}t  + j2\pi \frac{ldf_{c}\sin\theta_p}{c}} + w\left(t\right).
			\vspace{-0.1cm}
			\label{eqn:OFDM2}
			\end{align}
			\fi	
		}	
		\ac{ofdm} radar processing is based on 
		matched filtering \cite{Garmatyuk2010Multifunctional}. Its performance is determined by the complex weights, which can be optimally designed according to some requirements, e.g., the maximum peak-to-average ratio of the transmit signal\cite{Huang2015Low}, or the Cram{\'e}r-Rao bound  \cite{Turlapaty2014Range}.
		\label{Box:ofdmradar}
	\end{tcolorbox}	
	
	\vspace{-0.2cm}
	\subsection{Communications Waveform-Based Schemes}
	\label{subsec:jrc_Waveform}
	\vspace{-0.1cm}
	Another common \ac{dfrc} strategy is to utilize standard communications signals for probing, as illustrated in the upper-right subfigure in Fig. \ref{fig:DFRCCompare}.
	The majority of communications waveform-based designs in the literature utilize \ac{ofdm} signalling,  especially for automotive applications. 
	In the sequel, we first briefly review spread spectrum based \ac{dfrc} systems, followed by a more detailed presentation of shared \ac{ofdm} waveforms, \Revise{and a description how structured vehicular communications protocols can be used for sensing}. 

	\subsubsection{Spread Spectrum Waveforms}
	Spread spectrum techniques transmit a communications signal with a given bandwidth over a much larger spectral band, typically using spread coding or frequency hopping.  The usage of spread spectrum signals for radar probing was studied in  \cite{Sturm2011Waveform}.  
	The main drawback of spread spectrum  \ac{dfrc} design is that the radar dynamic range is limited, which is a byproduct of the imperfect auto-correlation properties of the spreading sequences\cite{Sturm2011Waveform}. In addition, accurately recovering the target velocity from spread spectrum echoes is typically computationally complex, limiting the applicability of such \ac{dfrc} systems. Finally, 
	high speed \acp{adc} is required for wideband spectrum spread waveforms, as de-chirp used in \ac{fmcw} is not applicable, increasing cost and complexity.
	
	\subsubsection{\ac{ofdm} Waveforms}
	\label{subsubsec:Sharedofdm}
	The most common communications waveform-based approach is to utilize \ac{ofdm} signalling. 
	\ac{ofdm} is a popular digital communications scheme due to its spectral efficiency, inherent ability to handle inter-symbol interference, and the fact that it can be implemented using relatively simple hardware components\cite{Hwang2009OFDM}. Since first proposed in \cite{levanon2000multifrequency}, \ac{ofdm} has received extensive attention as a radar waveform, especially for automotive radar, due to its high flexibility, adaptability in transmission, and since, unlike \ac{fmcw}, it does not suffer from  range-Doppler coupling  \cite{Franken2006Doppler}. The fact that \ac{ofdm} is commonly utilized in both radar and communications indicates its potential for \ac{dfrc} systems. 
	
	\Revise{Compared with case where  the coefficients $\{a_{m,n}\}$ in the \ac{ofdm} waveform are specifically designed for radar  (see \emph{\ac{ofdm} Waveform Radar} on Page \pageref{Box:ofdmradar}), the complex weights of the dual-function \ac{ofdm} waveform are  the communications symbols. The setting of the waveform parameters
		can have a notable effect on each functionality. 
		The work \cite{Braun2009Parametrization} designed the sub-carrier spacing according to the maximum unambiguous range and the maximum velocity. 
		In \cite{Liu2017Adaptive}  channel knowledge was used to allocate power  between the subcarriers to maximize the sum of the  data rate and the mutual information between the received echoes and the target impulse response.}
	\label{txt:OFDMProcees}
	\Revise{Radar processing of \ac{ofdm} waveforms utilizes  matched filtering, which depends on the transmitted data, causing high level sidelobes.
		This data dependency can be eliminated by dividing each subcarrier by its corresponding symbol \cite{Sturm2011Waveform}.	The range and velocity of each target are then estimated using a two dimensional \ac{dft} in the carrier domain and slow time domain (between different symbols), respectively.}

	\Revise{\ac{ofdm} can be naturally combined with \ac{mimo} radar which transmits orthogonal waveforms from each antenna (see \emph{\ac{mimo} Radar \label{Revise:OFDM-MIMO}} on Page \pageref{Box:mimoradar}) by assigning different subcarriers to different transmit elements.} Several works have studied how to divide the subcarriers among the elements. The proposed methods include division by
	equidistant sub-carrier interleaving \cite{Sturm2013Spectrally}; non-equidistant subcarrier interleaving\cite{Hakobyan2016A}; and random assignments \cite{Knill2019Random}.

	

	A drawback of using shared \ac{ofdm} waveforms in vehicular systems stems from the fact, when utilized from moving vehicles, \ac{ofdm} exhibits subcarrier misalignment, degrading the maximal radar unambiguous range \cite{Franken2006Doppler}.
	Additional drawbacks are related to  hardware constraints: 
	Wideband \ac{ofdm} waveforms require high rate \acp{adc}, affecting the system cost and power consumption. Another hardware limitation of \ac{ofdm} compared to monotone waveforms is its high peak-to-average power ratio, inducing distortion in the presence of practical non-linear amplifiers.
	A weighted \ac{ofdm} method was proposed to control the maximum peak-to-average power ratio\cite{Huang2015Low,Turlapaty2014Range}.
	In order to utilize \ac{ofdm} with narrowband transmissions, one can apply stepped frequency methods, as proposed in 
	\cite{Lellouch2015Stepped}.

	\begin{tcolorbox}[float,
		toprule = 0mm,
		bottomrule = 0mm,
		leftrule = 0mm,
		rightrule = 0mm,
		arc = 0mm,
		colframe = myblue,
		colback = mypurple,
		fonttitle = \sffamily\bfseries\large,
		title ={\ac{mimo} Radar}]	
		\ac{mimo} radar uses multiple 
		transmit and receive antennas. By transmitting orthogonal waveforms from each antenna, one can generate a virtual array with larger aperture, increasing the angular resolution  without requiring additional hardware elements. \Revise{While \ac{mimo} radar can also be combined with non-orthogonal waveforms, we focus on such systems transmitting orthogonal waveforms, which is the common practice in \ac{mimo} radar~\cite{Bliss2003The}}.
		
		To formulate \ac{mimo} radar transmission, let $L_T$ and $L_R$ be the numbers of transmit and receive antenna elements, respectively. The adjacent distances of the transmit antenna and the receive antenna are $d_T$ and $d_R$, respectively. A common practice is to set $d_T = L_R d_R$.
		We use $\myVec{s}\left(t\right) = \left[s_0\left(t\right), s_1\left(t\right), \cdots, s_{L_T - 1}\left(t\right)\right]^{T}$ for the transmit waveforms, which are orthogonal, namely, $\int s_{l}\left(t\right)s_{l^{'}}^{*}\left(t\right)dt = \delta\left(l-l^{'}\right)$, where $\delta\left(\cdot\right)$ is the Kronecker delta. For simplicity, we consider targets associated with a particular range and Doppler bin. The received signal is given by
		\vspace{-0.2cm}
		\begin{equation}		
		\myVec{y}\left(t\right) = \sum_{p=0}^{P-1}\alpha_{p} \myVec{a}^{T}\left(\theta_p\right)\myVec{s}\left(t\right)\myVec{b}\left(\theta_p\right) + \myVec{w}\left(t\right),   
		\label{eqn:MIMORx1}
		\vspace{-0.2cm}
		\end{equation}
		where $\myVec{a}\left(\theta\right) := \big[1, e^{j2\pi f_c{d_T\sin\theta}/{c}}, \cdots, e^{j2\pi f_c {\left(L_T - 1\right)d_T\sin\theta}/{c}}\big]^{T}$ is the transmit steering vector,  $\myVec{b}\left(\theta\right) := \big[1, e^{j2\pi f_c{d_R\sin\theta}/{c}}, \cdots, e^{j2\pi f_c {\left(L_R - 1\right)d_R\sin\theta}/{c}}\big]^{T}$ is the receive steering vector in direction $\theta$ and $\myVec{w}\left(t\right)$ is white Gaussian noise. Applying matched filtering and vectorization yields
		\ifsingle		
		\vspace{-0.2cm}
		\begin{align}
		\tilde{\myVec{y}} = {\rm vec}\left(\int \myVec{y}\left(t\right)\myVec{s}^{H}\left(t\right)dt\right) = \sum_{p=0}^{P-1} \alpha_p \myVec{a}\left(\theta_p\right)\otimes \myVec{b}\left(\theta_p\right) + \tilde{\myVec{w}},
		\vspace{-0.2cm}
		\label{eqn:MIMORx2}
		\end{align}
		\else
		\begin{align}
		\tilde{\myVec{y}} &= {\rm vec}\left(\int \myVec{y}\left(t\right)\myVec{s}^{H}\left(t\right)dt\right) \notag \\
		&= \sum_{p=0}^{P-1} \alpha_p \myVec{a}\left(\theta_p\right)\otimes \myVec{b}\left(\theta_p\right) + \tilde{\myVec{w}},
		\vspace{-0.2cm}
		\label{eqn:MIMORx2}
		\end{align}
		\fi		
		where ${\rm vec}\left(\cdot\right)$ is the vectorization operator, $\tilde{\myVec{w}} := {\rm vec} \left(\int \myVec{w}\left(t\right)\myVec{s}^{H}\left(t\right)\right)$, and $\otimes$ is the Kronecker
		product. Since $\myVec{a}\left(\theta_p\right)\otimes \myVec{b}\left(\theta_p\right) = \left[1, e^{j2\pi f_c {d_R\sin \theta_p}/{c}}, \cdots, e^{j 2 \pi f_c {\left(L_{T}L_{R} - 1\right)d_R \sin \theta_p}/{c}}\right]^{T}$, it holds that \ac{mimo} radar achieves an equivalent angle resolution of a phased array radar with $L_T L_R$ receive antennas in this configuration, effectively enhancing the angular resolution by a factor of $L_T$.
		
		\label{Box:mimoradar}
	\end{tcolorbox}	

	\subsubsection{Protocol-Oriented \ac{dfrc} Methods}
	\label{subsec:jrc_Protocol}
	An alternative strategy is to exploit the existing communications protocols, utilizing them as an automotive radar waveform. 
	Here, there is no compromise in the communications part, and the radar functionality is a byproduct of the  protocol, which is typically  IEEE 802.11p or IEEE 802.11ad \cite{Reichardt2012Demonstating,Kumari2015Investigating,Kumari2018IEEE80211ad,Muns2017Beam,Kumari2018Virtual,Grossi2017Opportunistic}.   	
	The IEEE 802.11p standard focuses on vehicular communications, and supports short range device-to-device transmissions for safety applications. This   protocol operates in the $5.9\  \mathrm{GHz}$ band and uses \ac{ofdm} signaling. Consequently, its transmissions  realize a \ac{dfrc} system with an \ac{ofdm} shared waveform, as proposed in~\cite{Reichardt2012Demonstating}. 
	
	IEEE 802.11ad is a generic standard for short range \ac{mmwave} communications operating at $60\  \mathrm{GHz}$. Its large bandwidth  enables higher data rates for communications, and better accuracy/resolution for radar operation. In order to avoid the usage of data-dependent waveforms, it has been proposed to utilize the a-priori known IEEE 802.11ad  preamble for radar probing  \cite{Kumari2015Investigating,Kumari2018IEEE80211ad,Muns2017Beam}. 
	As the preamble now affects radar performance, the work \cite{Kumari2018Virtual} studied the design of radar suitable preamble sequences. In such \ac{mmwave} communications, highly directional beams are used. Once the communications data link is established, radar can only reliably detect  targets located in the assigned beam direction.
	Several approaches have been proposed to extend the scanning area at the cost of power reduction in \cite{Grossi2017Opportunistic}. 
	
	The main benefit of protocol-based \ac{dfrc} designs is that they implement radar  with minimal effect on the communications functionality. As such, its radar capabilities are quite limited.  
	The radar coverage area is restricted by the directionally beamformed \ac{mmwave} transmission.  
	\label{txt:DutyCycle}
	\Revise{In  addition  to  its  restricted coverage  area,  the  scheme  has  a  relatively  low  radar  duty  cycle  as  only  the  preamble  is  utilized  for probing, limiting its detection range in vehicular systems operating under peak power constraints.
		\label{Protol_Conclusion}}
	


	\smallskip
	\label{txt:JRCWaveConc}
	To conclude,  communications waveform-based \ac{dfrc} approaches, and particularly using shared \ac{ofdm} signalling, enable transmitting high data rates by utilizing conventional digital communications schemes. The fact that \ac{ofdm} is widely studied for both radar and communications makes it an attractive \ac{dfrc} design. \Revise{In the context of autonomous vehicles, several drawbacks must be accounted for: First, in order to radiate enough power on the target, radar waveforms are typically beamformed to be directional. The communications receivers should thus be located in the radar beam in order to observe high signal-to-noise ratios. Such transmissions may thus be more suitable to serve as a secondary communications channel, in addition to a possible cellular-based \ac{v2x} technology which can communicate with the receivers in the omnidirection. Similarly, protocol-oriented schemes, which utilize standard communications transmission while exploiting its structure for probing, is more likely to provide additional sensing capabilities to a dedicated automotive radar. 
		Finally, 
		relatively costly hardware components are required for generating wideband waveforms and sampling their reflections. Despite these drawbacks, sensing using communications waveforms is considered to be a promising \ac{dfrc} approach for autonomous vehicles \cite{Patole2017}.}   
	
	

	\vspace{-0.2cm}
	\subsection{Radar Waveform-Based Techniques}
	\label{subsec:jrc_index}
	\vspace{-0.1cm}
	\ac{dfrc} systems can also  be designed by embedding the communication message in conventional radar waveforms, as illustrated in the bottom-right subfigure in Fig. \ref{fig:DFRCCompare}. These techniques are divided into two categories: The first approach modifies the radar waveform to incorporate digital modulations; the second method utilizes \ac{im}, conveying data bits via the indices of certain radar parameters. 
	
	\subsubsection{Modified Radar Waveforms} A possible approach to embed digital communications  into an existing radar system is to modify the waveform to include  modulated symbols. For example, the traditional \ac{fmcw} (see \emph{\ac{fmcw} Radar} on Page \pageref{Box:fmcw}) can be modified to include phase-modulated symbols by replacing the $m$th pulse $s_{m}(t)$, defined in the box on Page \pageref{Box:fmcw}, with $s_{m}(t)e^{j\phi_m}$, where $\phi_m$ encapsulates the information message in the form of, e.g.,  \acl{cpm} as proposed in \cite{Sahin2017Anovel}. Alternatively, the linear frequency of the pulse can convey information via frequency modulation \cite{Saddik2007Ultra}, for example, by using a positive frequency modulation rate $\gamma$ to transmit the bit $1$ and a negative value for $0$.
	While these schemes are typically power efficient \cite{Sturm2011Waveform} and with low complexity, their communication rate is very limited.
	
	Higher communication rates can be obtained by utilizing multiple orthogonal waveforms and beamforming. Assume $J$ orthogonal waveforms $\left\{s_j\left(t\right) \right\}_{j=0}^{J-1}$ are simultaneously transmitted from an antenna array, and let $\{\myVec{u}_{j}\}_{j=0}^{J-1}$ be the corresponding beamforming vectors. The transmit signal is expressed as $\myVec{s}\left(t\right) = \sum_{j=0}^{J-1}\myVec{u}_{j}s_j\left(t\right)$. In the communications receiver, the received signal is $y^{(c)}\left(t\right) = \myVec{g}_c^{T} \myVec{s}\left(t\right) + w^{(c)}\left(t\right)$, where $\myVec{g}_c$ and $w^{(c)}\left(t\right)$ are the channel response and additive noise, respectively. By applying matched filtering with the orthogonal waveforms, the receiver obtains the vector $\myVec{y}^{(c)} = \big[y_0^{(c)}, y_1^{(c)}, \cdots, y_{J-1}^{(c)}\big]^{T}$, where $y_j^{(c)} =  \myVec{g}_c^{T}\myVec{u}_{j} + w_j^{(c)}\left(t\right)$. The communication data bits can be conveyed by modulating the amplitude\cite{Hassanien2016Dual} or phase\cite{Hassanien2016Non} of $\myVec{g}_c^{T}\myVec{u}_{j}$. Although the communication rates are improved by transmitting multiple waveforms, the system complexity is also increased, and the transmitter must have a-priori knowledge of the communications channel $\myVec{g}_c$. Furthermore, it is difficult to guarantee that the envelope of the transmit signal is constant modulus, which may reduce power efficiency in transmission.

	

	\subsubsection{\ac{im}-Based Techniques}
	\ac{im} is a promising communications technique, gaining growing interest due to its high energy and spectral efficiency\cite{Basar2017Index}. 
	Instead of using conventional modulations, \ac{im}  embeds data bits into the indices of certain transmission building blocks\cite{Basar2017Index}. 
	These building blocks, including spatial allocation and frequency division, are also important waveform parameters for radar. \ac{im}-based \ac{dfrc} techniques thus embed the digital message into the combination of radar waveform parameters. The term \textit{index} represents the radar parameters, such as carrier frequency, time slot, antenna allocation, or orthogonal waveforms in \Revise{\ac{mimo} radar with orthogonal waveforms}.  Consequently, such \ac{dfrc} systems use unmodified conventional radar schemes, and the ability to communicate is encapsulated in the parameters of the transmission. 
	While \ac{im}-based \ac{dfrc} schemes are the focus of ongoing research, existing methods typically build upon either {\ac{mimo} radar} or \ac{far} schemes. \Revise{While \ac{mimo} radar can in general utilize orthogonal or non-orthogonal waveforms,  we henceforth use the term \emph{\ac{mimo} radar} for such schemes utilizing orthogonal waveforms, which is the typical approach in \ac{mimo} radar\cite{Bliss2003The} .} 
	
	\begin{tcolorbox}[float,
		toprule = 0mm,
		bottomrule = 0mm,
		leftrule = 0mm,
		rightrule = 0mm,
		arc = 0mm,
		colframe = myblue,
		colback = mypurple,
		fonttitle = \sffamily\bfseries\large,
		title =Frequency Agile Radar]	
		
		A promising approach to tackle mutual interference between radars is to utilize \ac{far} \cite{Axelsson2007}. Here, a sub-band waveform (of a much narrower bandwidth compared to the available band) is transmitted in each cycle, and its central frequency varies randomly from cycle to cycle. These random variations reduce the spectral collision probability from neighbouring radars.
		
		\ \ To formulate the signal model, we use $\mathcal{F} = \left\{f_c + n \Delta f| n = 0, 1, \cdots, N-1\right\}$ to denote carrier frequency set, where $\Delta f$ is the carrier spacing. During the $m$th transmit pulse, the transmitted signal is $s_m\left(t\right)= e^{j2\pi f_m t}$, where $f_m$ is randomly chosen from $\mathcal{F}$. After demodulation, the received signal at the $l$th receiving antenna can be expressed using the notations of \eqref{eqn:FMCWrx} as 
		\textcolor{black}{		
			\ifsingle
			\vspace{-0.2cm}
			\begin{align}
			y_{m,l}\left(t\right) = \sum_{p=0}^{P-1}\alpha_{p} e^{-j 2 \pi f_m \frac{2r_p}{c} - j 2\pi f_m \frac{2v_p m T_{PRI}}{c} + j2\pi \frac{ldf_{m}\sin\theta_p}{c}} +w\left(t\right).
			\label{equ:FARSignal}
			\vspace{-0.2cm}
			\end{align}
			\else
			\begin{align}
			y_{m,l}\left(t\right) &= \sum_{p=0}^{P-1}\alpha_{p} e^{-j 2 \pi f_m \frac{2r_p}{c} - j 2\pi f_m \frac{2v_p m T_{PRI}}{c}}\notag \\
			& \times e^{j2\pi \frac{ldf_{m}\sin\theta_p}{c}} +w\left(t\right).
			\label{equ:FARSignal}
			\vspace{-0.2cm}
			\end{align}
			\fi
		}
		\ \ Using matched filtering, \ac{far} can synthesize a large bandwidth and enables to generate high range resolution profiles. However, the random changing of carrier frequency leads to a high sidelobe level, which affects the detection of weak targets. To mitigate the sidelobe problem, \acl{cs} methods can be applied for range-Doppler processing\cite{Huang2014TAES}, while recovery guarantees for such methods are provided in \cite{Huang2018Analysis,Wang2019Analysis} under sparse and block-sparse target scenes, respectively.  
		\label{Box:far}
	\end{tcolorbox}		
	
	\emph{\ac{im} for \ac{mimo} radar:} 
	\label{txt:ImMimoRadar}
	The work \cite{BouDaher2016Towards} proposed to combine \ac{mimo} radar with \ac{im} by embedding the bits in the assignment of the orthogonal waveforms across the transmit antennas. For a \ac{mimo} radar with $L_T$ transmitting antennas, there are $L_T!$ possible arrangements  in each \ac{pri}, supporting a maximal rate of $\log L_T !$ bits per \ac{pri}. In \cite{Wang2019Dual}, this approach was extended to sparse array \ac{mimo} radar configurations,  where only $K$ out of $L_T$ transmit elements are active in each \ac{pri}. As a consequence, it requires only $K$ transmit orthogonal waveforms, represented  (with a slight notation abuse) by the vector $\myMat{s}\left(t\right) = \left[s_0\left(t\right), s_1\left(t\right), \cdots, s_{K-1}\left(t\right)\right]^T$. The transmitted $L_T \times 1$ vector $\tilde{\myVec{s}}\left(m, t\right)$ in the $m$th \ac{pri} is a  permutation of $\myMat{s}\left(t\right)$, i.e., it is given by  $	\tilde{\myVec{s}}\left(m,t\right) =  \myMat{\Omega}^{T}_{\mathrm{M}}\myMat{\Lambda}^{T}\left(m\right)\myVec{s}\left(t\right)$, where $\myMat{\Lambda}\left(m\right)$ is a $K \times K$ permutation matrix,  and   $\myMat{\Omega}_{\rm M}\left(m\right)\in \left\{0, 1\right\}^{K\times L_{T}}$ is the antenna selection matrix which has a single non-zero entry in each row. 
	When the channel is memoryless, the  signal received at the communications receiver 
	is 
	\vspace{-0.2cm}
	\begin{equation}
	y^{\left(c\right)}\left(m,t\right) =  \myVec{g}_c^{T}\left(m\right)\tilde{\myVec{s}}\left(m,t\right) + w^{(c)}\left(m,t\right),
	\label{eqn:MIMO_IM1}
	\vspace{-0.2cm}
	\end{equation} 
	where  $\myVec{g}_c$ is the $L_T \times 1$  channel vector, and $ w^{(c)}\left(m,t\right)$ is the additive noise. After matched filtering with the orthogonal waveforms, the obtained vector can be written as
	\ifsingle	
	\vspace{-0.2cm}
	\begin{align}
	\myVec{y}^{\left(c\right)}\left(m\right) = \int y^{\left(c\right)}\left(t, m\right) \myVec{s}\left(t\right)dt =   \myMat{\Lambda}\left(m\right)\myMat{\Omega}_{\rm M}\left(m\right)\myVec{g}_c + \myVec{w}\left(m\right).
	\vspace{-0.2cm}
	\label{eqn:MIMO_IM2}
	\end{align}	
	\else
	\begin{align}
	\myVec{y}^{\left(c\right)}\left(m\right) &= \int y^{\left(c\right)}\left(t, m\right) \myVec{s}\left(t\right)dt \notag \\&=   \myMat{\Lambda}\left(m\right)\myMat{\Omega}_{\rm M}\left(m\right)\myVec{g}_c + \myVec{w}\left(m\right).
	\vspace{-0.2cm}
	\label{eqn:MIMO_IM2}
	\end{align}	
	\fi	
	The communication message can be embedded in $\myMat{\Lambda}\left(m\right)\myMat{\Omega}_{\rm M}\left(m\right)$ in \eqref{eqn:MIMO_IM2}, i.e., the product of permutation matrix and the selection matrix. As there are $\binom{L_T}{K}$ kinds of antenna selection pattern and $K!$ kinds of waveform permutations, up to $\log_2{\binom{L_T}{K}}+\log_2{K!}$ bits can be encapsulated in each \ac{pri}. 
	
	

	\emph{\ac{im} via \ac{far}:} 
	\ac{far} (see \emph{Frequency Agile Radar}  on Page \pageref{Box:far}) is a radar scheme designed for  congested environments. The carrier frequencies of \ac{far} change randomly from pulse to pulse, allowing to achieve an ergodical wideband coverage, while utilizing narrowband waveforms and enabling to mitigate interference from neighbouring radars. 
	The work \cite{Wang2019Codesign} proposed a \ac{dfrc} system which embeds a digital message into the  permutation of the agile carrier frequencies. For a carrier set with $N$ different carrier frequencies, there are $N!$ different carrier frequency permutations that can be utilized for information embedding.

\begin{figure}	
		\centerline{\includegraphics[width=\columnwidth] {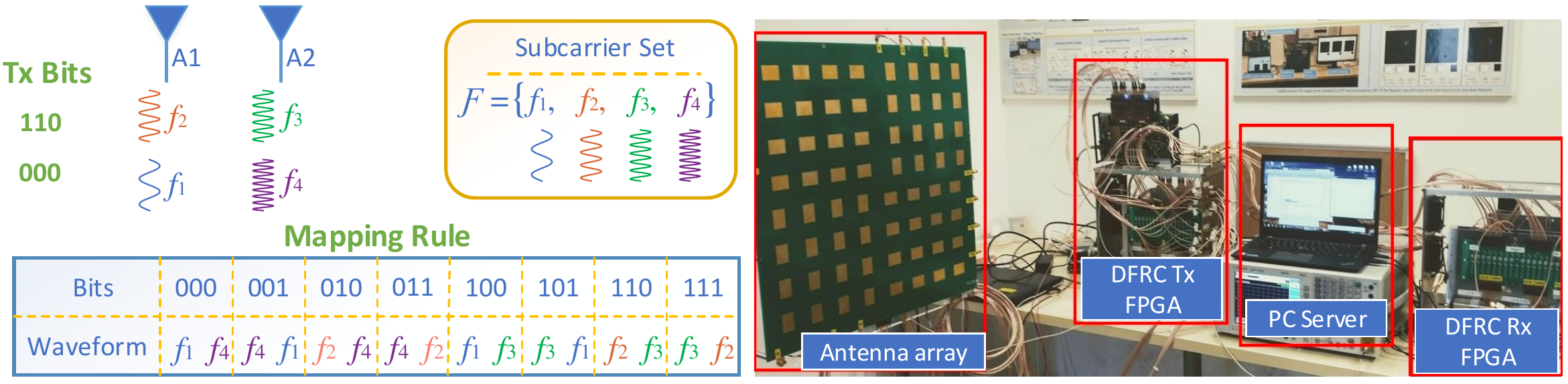}}
		\vspace{-0.2cm}
		\caption{An illustration of \ac{im}-\ac{far} \cite{Huang2019multi} (left) and a hardware prototype equipped with $64$ antenna elements (right) which was demonstrated in 2019 IEEE ICASSP. In the example  (left), the array consists of $L_T = 2$  elements, divided into $K = 2$ sub-arrays of $L_K = 1$ elements.  The carrier set is $\mathcal{F} = \left\{f_1, f_2, f_3, f_4\right\}$. The mapping rule represents the codebook. }
		\label{fig:IMFAR}
	\end{figure}

	In \cite{Huang2019multi,Huang2019A}, a \ac{dfrc} system is proposed based on multi-carrier agile waveforms and \ac{im}. Unlike traditional \ac{far}, here multiple carriers are simultaneously sent from several sub-arrays of transmit antennas. 
	For a \ac{dfrc} system with $L_T$ transmit antenna elements and a possible carrier frequency set $\mathcal{F}$ of cardinality $N$, the corresponding information embedding consists of two stages: In the $m$th pulse, $K < N$ carriers, denoted by the set $\left\{f_{m,0},\cdots,f_{m,K-1} \right\}$, are first selected from $\mathcal{F}$.  Then, the  antenna array is divided into $K$ sub-arrays, where each sub-array has $L_K=\frac{L_T}{K}$ elements.  
	The transmitted signal of the multi-carrier frequency agile \ac{dfrc} system in the $m$th \ac{pri} is expressed as 
	\ifsingle	
	\vspace{-0.2cm}	
	\begin{equation}
	\myVec{\tilde{s}}\left(m,t\right) = \sum_{k=0}^{K-1}\myMat{\Omega}_{\rm F}\left(m,k\right)\myVec{u}\left(\theta, f_{m,k}\right){\rm rect}\left(\frac{t - mT_p}{T_p}\right)e^{j2\pi f_{m,k}\left(t - mT_p\right)},
	\vspace{-0.2cm}
	\end{equation}
	\else	
	\begin{align}
	\myVec{\tilde{s}}\left(m,t\right) = \sum_{k=0}^{K-1}&\myMat{\Omega}_{\rm F}\left(m,k\right)\myVec{u}\left(\theta, f_{m,k}\right){\rm rect}\left(\frac{t - mT_p}{T_p}\right)\notag \\
	&\times e^{j2\pi f_{m,k}\left(t - mT_p\right)},
	\vspace{-0.2cm}
	\end{align}
	\fi	
	where $\theta$ is the beam steered direction, $\myVec{u}\left(\theta, f_{m,k}\right)$ is the radar beamforming vector for the $k$th carrier with frequency $f_{m,k}$, and $\myMat{\Omega}_{\rm F}(m,k)$ is the selection matrix which determines the transmit antennas of carrier with frequency $f_{m,k}$. The communication message is embedded into the antenna allocation pattern as well as the selection of carrier frequencies.
	The number of antenna allocation patterns is $\frac{L_T!}{{\left(L_K!\right)}^{K}}$, and there are $\binom{N}{K}$ possible combinations of carrier selections. Hence, the total number of transmission patterns that can be used for information embedding is $\binom{N}{K}\cdot\frac{L_T!}{{\left(L_K!\right)}^{K}}$. An illustration of this scheme, as well as a hardware prototype designed to demonstrate its feasibility, are shown in Fig. \ref{fig:IMFAR}.

	\label{txt:IMDecode}
	\Revise{Since \ac{im}-based \ac{dfrc} systems utilize conventional radar waveforms, radar detection is carried out using standard methods. For example, \ac{far} detection is based on matched filtering followed by  \acl{cs} recovery \cite{eldar2012compressed}. Symbol detection at the communications receiver can be realized using the maximum likelihood rule, or alternatively, via a reduced complexity \ac{im} detector, see, e.g., \cite{Huang2019A}.}

	
	\smallskip
	\label{txt:discIM}
	The main advantage of radar waveforms-based \ac{dfrc} methods is that they provide the ability to communicate with minimal degradation to the performance of the radar scheme from which the technique originates. For example, the radar performance of \ac{mimo} radar as well as \ac{far} combined with \ac{im} are roughly equivalent to their radar-only counterparts \cite{Huang2019multi}, respectively. In particular,  \ac{far} is  attractive for automotive radar  due to its inherent applicability in congested setups and compliance with simplified hardware. Nonetheless, the communications functionality of radar waveform-based \ac{dfrc} systems is relatively limited in throughput, and typically results in increased decoding complexity, making it more suitable to serve as an alternative channel in addition to existing, e.g., cellular-based, vehicular communications, rather than replacing the latter.   
	
	\vspace{-0.2cm}
	\subsection{\Revise{Joint Waveform Design }}
	\label{subsec:jrc_jointwaveform}
	\vspace{-0.1cm}
	\Revise{The approaches detailed so far are all based on traditional radar and/or communications signalling. A \ac{dfrc} system is then obtained by either designing the conventional waveforms to coexist, as detailed in Subsection \ref{subsec:jrc_Coexistence}, or alternatively, by using only one standard waveform while extending it to be dual-functional. Using traditional signalling has clear advantages due to their established performance and applicability with existing hardware devices. Nonetheless, the fact that these waveforms were not originally designed for \ac{dfrc} scenarios implies that one can achieve improvement by deriving dedicated dual-function waveforms, as illustrated in the bottom-left subfigure of Fig. \ref{fig:DFRCCompare}. }
	
	\Revise{Dedicated joint waveforms, which do not originate from conventional radar / communications  signalling, are designed according to a dual-function objective which accounts for the performance of both radar and communications  \cite{McCormick2017Simultaneous,Liu2018MUMIMO,Liu2018Toward}. Here, the transmitted joint signal is denoted by the ${L_{T}\times J}$ matrix $\myMat{X}$, where $J$ is the block length. We focus on a multi-user scenario with $L_U$ single antenna receivers. The signal received at the receivers and at the radar target with direction $\theta$ can be expressed as
		\vspace{-0.2cm}
		\begin{align}
		\myMat{Y}^{\left(c\right)} = \myMat{H}\myMat{U}\myMat{X} + \myMat{W}^{\left(c\right)},\  \mathrm{and} \  \myVec{y}^{\left(r\right)}_{\theta} = \myVec{a}^{T}\left(\theta\right)\myMat{U}\myMat{X},
		\label{equ:jointwaveform}
		\vspace{-0.2cm}
		\end{align}	
		where $\myMat{H}$ is an $L_{U}\times L_{T}$  channel matrix, $\myVec{U}$ is the joint beamformer, and $\myMat{W}^{\left(c\right)}$ is the additive noise term.}
	
	\Revise{Using formulation \eqref{equ:jointwaveform}, one can design the joint waveform $\myMat{X}$ in order to approach some desired observations at the communications receivers as well as the radar target, as proposed in \cite{McCormick2017Simultaneous}. 
		A possible drawback  is that  the signals received in other directions are not constrained, and thus the radar transmit beampattern may have a high sidelobe level outside the mainlobe. This can be overcome by restricting the radar beampattern\cite{Liu2018MUMIMO,Liu2018Toward}, which is in turn achieved by constraining the signal covariance. In particular, \cite{Liu2018MUMIMO} considered $\myMat{X}$ to be a communications signal and optimized the joint precoding to approach a pre-defined beampattern while meeting a minimal \ac{sinr} level at each  receiver. The work \cite{Liu2018Toward} designed the joint waveform  $\myMat{X}$  to minimize the multi-user interference under specific radar constraints, such as omindirectional or directional beampatterns, constant modulus designs, and waveform similarity.}

	\label{txt:discJoint}
	\Revise{Dual function waveforms specifically designed for \ac{dfrc} offer to balance radar and communications in a controllable manner. Furthermore, using  joint optimization, without being restricted to conventional waveforms, can potentially yield any achievable performance tradeoff between radar and communications. Despite these clear theoretical benefits, their application in automotive \ac{dfrc} system is still limited to date due to practical considerations. For example, current joint waveforms designs involve solving a relatively complex optimization problem, which depends on prior channel knowledge. In fast moving vehicles,  accurate instantaneous channel knowledge is difficult to obtain, and even when it is available, the optimization process must be frequently repeated, inducing increased computational burden.} 

	\vspace{-0.2cm}
	\subsection{Discussion}
	\label{subsec:jrc_Summary}
	\vspace{-0.1cm}
	The \ac{dfrc} methods surveyed here vary significantly in their characteristics such as radar performance, communication throughput,  complexity, and hardware requirements. Although several efforts have been made in the literature to characterize the achievable radar-communications trade-off in \ac{dfrc} systems \cite[Ch. VI]{Paul2017}, to date there is no unified joint measure which allows to rigorously evaluate different schemes. 
	
	\Revise{
		To demonstrate the challenge in comparing \ac{dfrc} methods, we numerically evaluate two promising schemes: \ac{ofdm} waveforms, which utilize communications signaling for radar probing, and the radar-based \ac{im} via \ac{far} method. In particular, we consider a single antenna  automotive radar in the $24$ GHz band divided into $1024$ bins, using the same configuration as in \cite{Sturm2011Waveform}. \ac{ofdm}  utilizes the complete frequency band,  while \ac{im}-\ac{far} uses a single subcarrier at each instances, embedding the message in its selected index, i.e., a total of $\log_2 1024 = 10$ bits per symbol. To guarantee that both methods operate with the same data rate, we group the \ac{ofdm} subcarriers into $10$ distinct blocks, and assign a binary phase shift keying symbol to each block.  Both schemes use the same pulse width, \ac{pri}, and power, 
		attempting to recover a point target with range $10$ [m] and relative velocity $5$ [m/s],
		while communicating over a Rayleigh flat fading channel. The resulting normalized \ac{mse} in target range recovery, as well as the communications \acl{ber}, are depicted in Fig. \ref{fig:RangeMSE}. Observing Fig. \ref{fig:RangeMSE} we note that \ac{ofdm}  achieves improved communications performance over \ac{im}-\ac{far}, while their radar performance is relatively similar. The results in Fig. \ref{fig:RangeMSE}, which are in favor of \ac{ofdm}-based \ac{dfrc} systems, are relevant for interference free scenarios, where a single \ac{dfrc} system probes the environment. In  dense scenarios with multiple  interfering devices, which model automotive systems in urban settings, \ac{far} is expected to be more capable of mitigating the mutual interference due to its random spectral sparsity~\cite{Huang2019multi}.}

	\ifsingle	
	\begin{figure}	
		\begin{minipage}[b]{0.5\columnwidth}
			\centering
			\centerline{\includegraphics[width= 0.9\columnwidth]{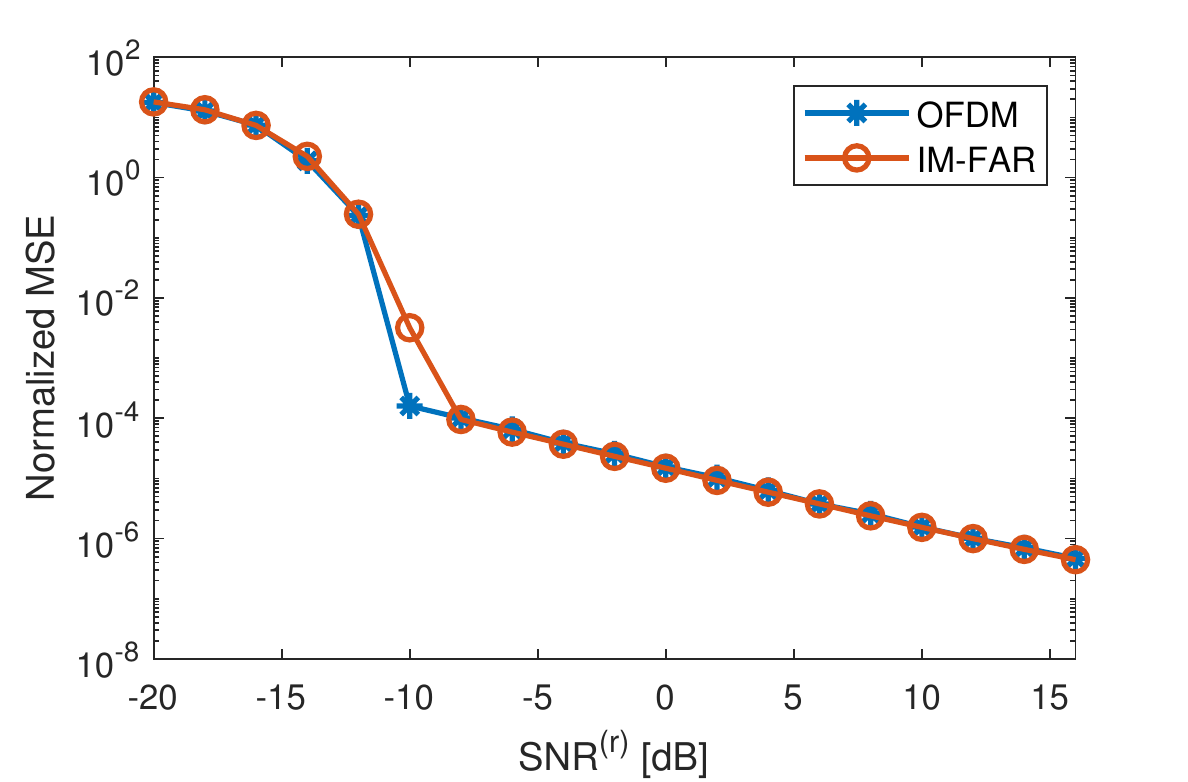}}
			\centerline{\small (a) Normalized \acs{mse} in range estimation}\medskip
		\end{minipage}
		\begin{minipage}[b]{0.5\columnwidth}
			\centering
			\centerline{\includegraphics[width= 0.9\columnwidth]{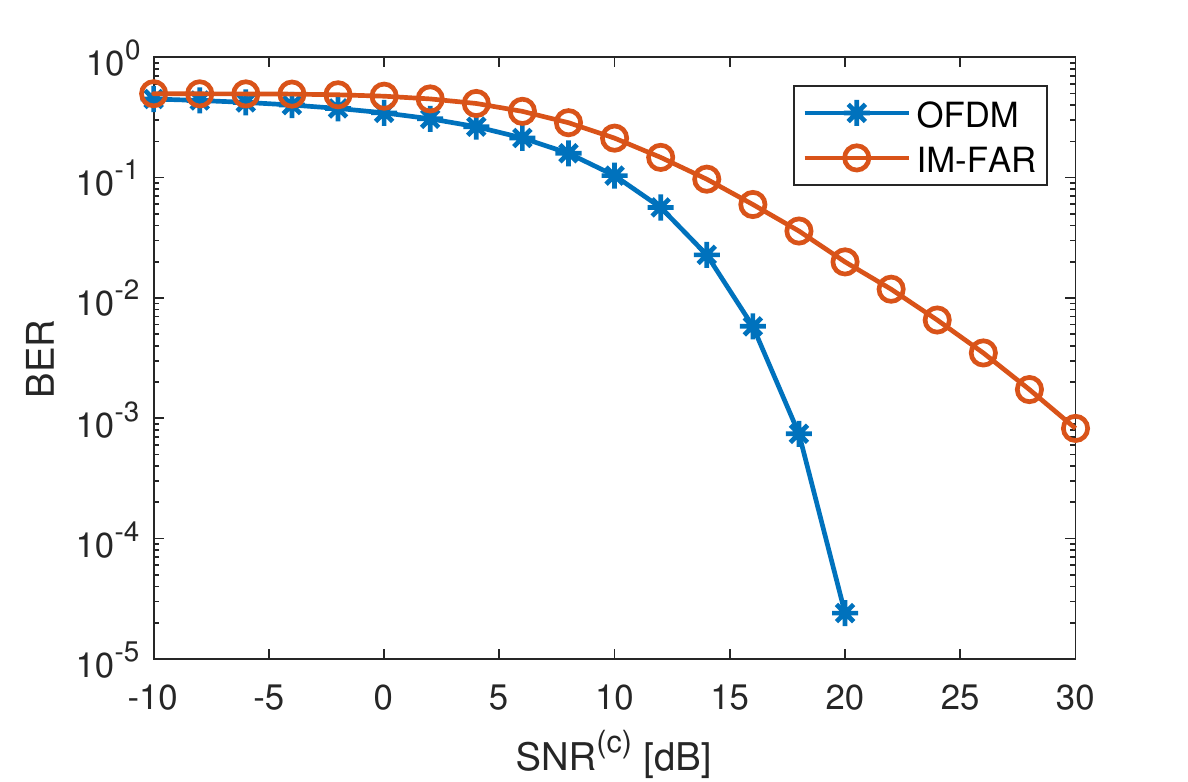}}
			\centerline{\small (b) Bit error rate}\medskip
		\end{minipage}
		\vspace{-1.0cm}
		\caption{Numerical comparison of \ac{ofdm}-based \ac{dfrc} systems to frequency agile radar with \ac{im}. }
		\label{fig:RangeMSE}
	\end{figure}
	\else
	\begin{figure}
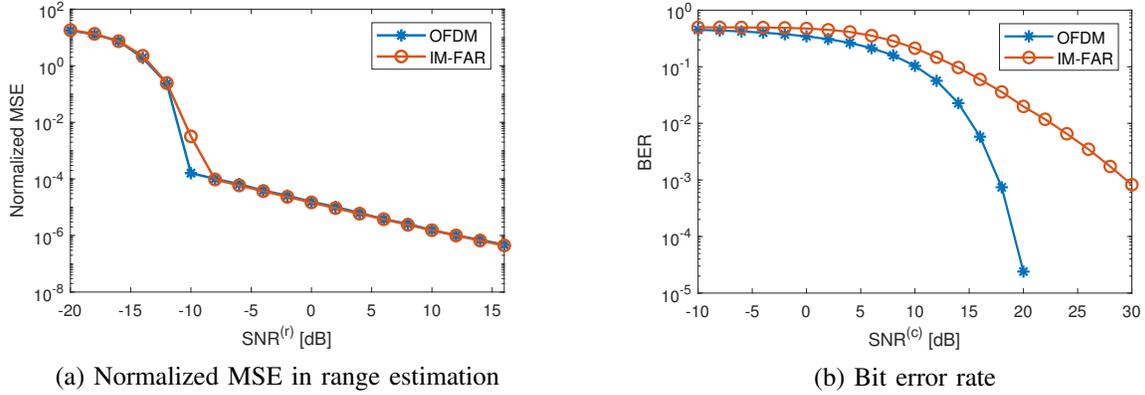
	
		\begin{minipage}[b]{\columnwidth}
			\centering
			\centerline{\includegraphics[width= 0.9\columnwidth]{fig/RadarMSE.eps}}
			\centerline{\small (a) Normalized \acs{mse} in range estimation}\medskip
		\end{minipage}
		\begin{minipage}[b]{\columnwidth}
			\centering
			\centerline{\includegraphics[width= 0.9\columnwidth]{fig/SimuBER.eps}}
			\centerline{\small (b) Bit error rate}\medskip
		\end{minipage}
		\vspace{-1.0cm}
		\caption{Numerical comparison of \ac{ofdm}-based \ac{dfrc} systems to frequency agile radar with \ac{im}. }
		\label{fig:RangeMSE}
	\end{figure}
	\fi

	Due to the difficulty in comparing \ac{dfrc} schemes, we schematically evaluate their radar vs. communications performance trade-off in Fig. \ref{fig:SchemComp}. 
	\Revise{Separate coordinated transmission methods}, which utilize individual signals for each functionality, support a broad range of possible performance combinations, determined by how the system resources are allocated between the functionalities. In particular, beamforming techniques, which require a-priori channel knowledge, allow the signals to utilize the full bandwidth and operation time, and thus have the potential to achieve improved performance compared to time/frequency division strategies. Nonetheless in the presence of multiple scatterers and communications receivers, which is the case in vehicular applications, obtaining accurate channel knowledge and mitigating mutual interference by beamforming may be infeasible, while spectral division can be applied with controllable complexity regardless of the number of receivers and their physical location.   
	
	Communications waveform-based approaches, particularly when using \ac{ofdm} transmission, support high data rates by utilizing conventional digital communications signals. Specifically, \ac{ofdm} is a digital communications scheme which has some of the characteristics of good radar waveforms. In the context of autonomous vehicles, a major limitation of this approach is that, since a single directed beam is used, the receiver should be located in the radar search area. Furthermore, \ac{ofdm} transmission requires relatively costly hardware,  and its radar capabilities are degraded when utilized by a moving vehicle. 
	
	
	Protocol-oriented approaches, being an extreme case of using a communications waveform for radar probing, offer to utilize existing vehicular communications protocols for sensing. They provide minimal communications degradation with limited radar capabilities. As such, these methods can be considered as an additional sensing technology, which should not replace dedicated automotive radar.
	
	Radar waveform-based schemes, especially \ac{im}-based \ac{dfrc} systems, can be naturally integrated into automotive radar systems with minimal effect to their performance. While \ac{mimo} radar implementing instantaneous wideband waveforms offers improved radar performance over frequency agile waveforms, the latter may be preferable to vehicular applications due to its robustness to congested environments and  reduced complexity. Nonetheless, the limited bit-rates of \ac{im} and its associated decoding complexity make such \ac{dfrc} schemes more suitable to provide an additional communications channel, independent of the cellular network.  The usage of such channels for safety and emergency messages can be valuable in autonomous vehicles, increasing the probability of their successful transmission.  
	
	\Revise{Joint waveform design techniques optimize a dual-function waveform in light of a combined constraints on each functionality. This joint approach has the potential of achieving any given tradeoff between radar and communications performances. Nonetheless, being a relatively new field of study, current dual function designs may not be suitable for automotive applications. In particular, current designs require instantaneous channel knowledge, limiting their application for self-driving vehicles. }
	
	To conclude, there is no single \ac{dfrc} method which is suitable for all scenarios and requirements encountered in autonomous vehicle applications. Understanding the advantages and disadvantages of each approach 
	will allow engineers to properly select the technologies incorporated into future self-driving cars.

	\ifsingle	
	\begin{figure}	
		\centerline{\includegraphics[width=0.45\columnwidth]{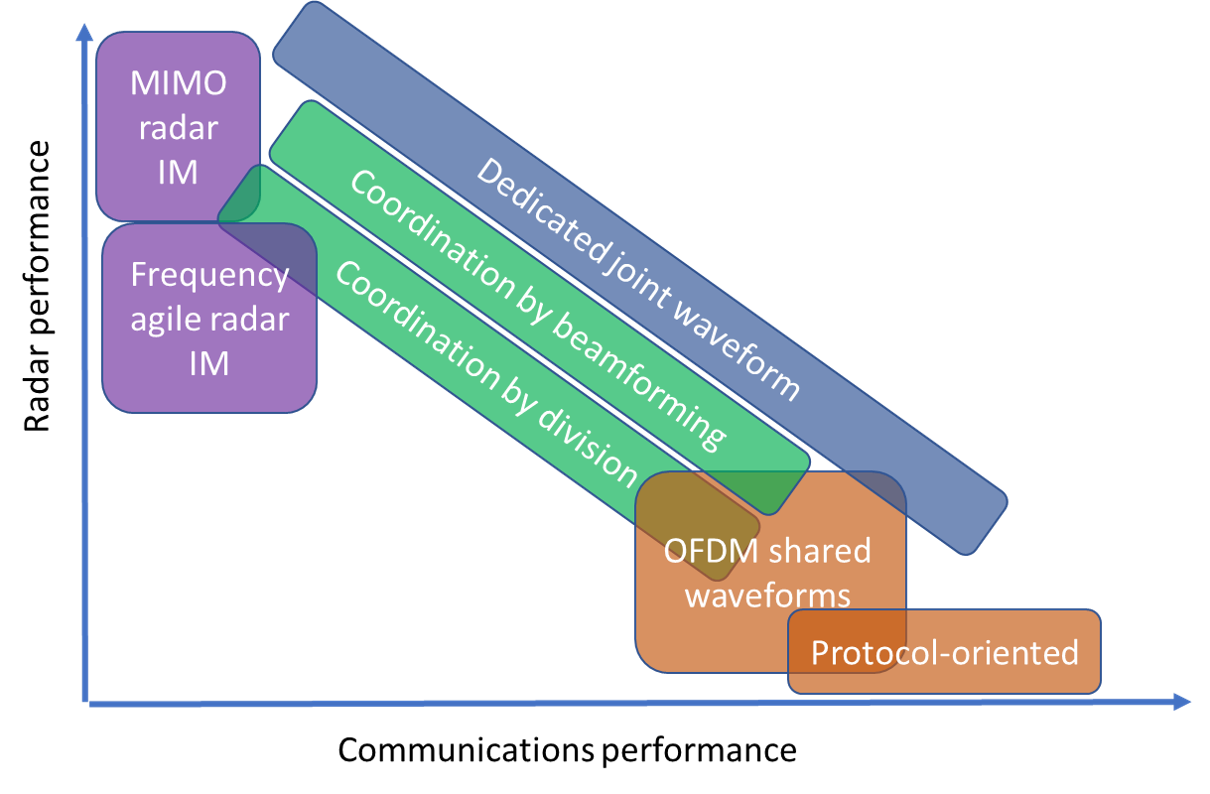}}
		\vspace{-0.35cm}
		\caption{Schematic comparison between considered \ac{dfrc} schemes in terms of their radar-communications trade-off. }
		\label{fig:SchemComp}
	\end{figure}	
	\else
	\begin{figure}	
		\centerline{\includegraphics[width=\columnwidth]{fig/SchemComp2.png}}
		\vspace{-0.35cm}
		\caption{Schematic comparison between considered \ac{dfrc} schemes in terms of their radar-communications trade-off. }
		\label{fig:SchemComp}
	\end{figure}
	\fi	

	\vspace{-0.2cm}	
	\section{Conclusions And Future Challenges}
	\label{sec:Conclusion}
	\vspace{-0.1cm}
	Autonomous vehicles implement wireless communications as well as automotive radar, which both require the transmission and reception of electromagnetic signals. Jointly designing these functionalities as \ac{dfrc} system provides potential gains in performance, size, cost, power consumption, and robustness, making it an attractive approach for autonomous vehicles. In this survey, we reviewed state-of-the-art in \ac{dfrc} designs focusing on their application for autonomous vehicles. To that aim, we first reviewed  the basics of automotive radars, briefly discussing associated radar waveforms such as \ac{fmcw}, frequency agile, and \ac{mimo} radar.  \textcolor{black}{The detailed intoductions of these waveforms and \ac{mimo} radar are given in the boxes acorss the paper.}  
	Then, we mapped existing \ac{dfrc} strategies, proposing their division into four main categories: Coexistence schemes which utilize independent waveforms for each functionality; communications waveform-based approaches where conventional communications signals are used for radar probing; radar waveform-based schemes which embed the digital message into standard radar technologies, and joint waveform design approaches which achieve the \ac{dfrc} system by deriving dedicated dual-function waveforms. The pros and cons of each category were analyzed according to the radar and communications requirements in vehicular scenarios. While we conclude that no single \ac{dfrc} scheme is suitable for all the scenarios in self-driving, our analysis can significantly facilitate the design of sensing and communications technologies for future autonomous vehicles. 
	
	While joint radar-communications designs have been studied for over a decade, they still give rise to a multitude of unexplored research directions, particularly in the context of autonomous vehicles. 
	On the theoretical side, the lack of a unified  performance measure makes it difficult to compare  approaches, and one must resort to heuristic arguments, as done in this article. Such an analysis will also uncover the fundamental limits of \ac{dfrc} designs, characterizing their optimal gain over well-studied separate systems. 
	From an algorithmic perspective, the utilization of joint non-standard radar and communications waveforms, utilized in some of the aforementioned strategies, can be facilitated by the development of dedicated recovery and decoding algorithms. For conventional waveforms, such as  \ac{ofdm} signals, efficient allocation of resources to optimize both functionalities is a relatively fresh area of study. 
	Additionally, the presence of multiple sensing vehicular technologies, such as vision-based sensing and \ac{lidar}, along with the ability to communicate with neighbouring devices which also sense their environment, give rise to potential improved understanding of the surroundings by properly combining these technologies.  
	Finally, on the practical side, future investigations are required to implement these strategies in  vehicular platforms and test their performance in real road environments. Such combined studies should allow to characterize the benefits and limitations of \ac{dfrc} systems for self-driving cars, allowing their theoretical potential to be translated into performance gains in this emerging and exciting technology.  
	
	\vspace{-0.2cm}

	\begin{spacing}{1.005}
		\bibliographystyle{IEEEtran}
		\bibliography{IEEEabrv,references}
	\end{spacing} 
	
\end{document}